# The Shift Towards Preprints in AI Policy Research: A Comparative Study of Preprint Trends in the U.S., Europe, and South Korea

Simon Suh
Department of Technology and Society
Stony Brook University

THIS PAPER HAS BEEN WITHDRAWN and Updated in the new link below
https://arxiv.org/pdf/2510.16477

**[Abstract]**

The adoption of open science has quickly changed the way artificial intelligence (AI) policy research is distributed globally. This study examines the regional trends in the citation of preprints, specifically focusing on the impact of two major disruptive events – the COVID-19 pandemic and the release of ChatGPT– on research dissemination patterns in the United States, Europe, and South Korea. Using bibliometric data from the Web of Science (2018-2024), this study tracks how global disruptive events influenced the adoption of preprints in AI policy research across different regions over time. By marking the timing of these disruptive events, the analysis reveals that while all regions experienced growth in preprint citations, the magnitude and trajectory of change varied significantly. The United States exhibited sharp, event-driven increases; Europe demonstrated steady acceleration; and South Korea maintained consistent, linear growth in preprint adoption. These findings suggest that global disruptions have accelerated preprint adoption, but regional factors critically shape how open science practices are integrated. This study emphasizes the need for future AI governance strategies to consider regional variability in research dissemination and highlights opportunities for further longitudinal and comparative research to deepen our understanding of open-access adoption in AI policy development.

## 1. Introduction

The field of artificial intelligence (AI) has experienced a revolutionary change, with new research being released at an incredible pace. This transition not only accelerated the development of AI technologies but also reshaped the way scientific information is communicated and authenticated. Within this landscape, AI policy research—which is scholarly work that informs the ethical, legal, and governance frameworks surrounding AI—plays a critical role in shaping responsible innovation and public trust. The credibility, timing, and accessibility of this research directly influence regulatory decisions, funding priorities, and international cooperation on AI standards.

Traditionally, peer-reviewed journals used to serve as the standard for academic credibility while acting as a filter to prevent low-quality research and ensuring the research validity, significance, and originality (Kelly et al., 2014). However, the AI research community has started to rely more on the use and reference of preprint platforms such as arXiv, which allow researchers to share their findings publicly before undergoing a formal peer review. This shift from traditional peer-reviewed publishing to preprint platforms raises critical questions about the shifting landscape of academic research on AI policy. While peer-reviewed journals remain crucial in maintaining the research quality and scholarly integrity, preprint platforms offer speed, accessibility, and immediate visibility—qualities that are valuable in the AI field where timely access to new knowledge is essential.

This paper investigates the increasing reliance on preprints in AI policy research and examines how two global disruptive events—the COVID-19 pandemic and the release of ChatGPT—have influenced this shift. Specifically, it addresses two key research questions: (1) What factors are driving the move from peer-reviewed journals to preprint platforms in AI policy research? (2) How do these trends vary across AI-leading countries such as the United States, Europe, and South Korea?

These regions were selected for their prominent roles in AI development, governance, and innovation. Each country has a unique institutional approach to open science and research dissemination, making them ideal for a comparative analysis of preprint adoption. By analyzing regional citation trends from 2018 to 2024, this study aims to provide a clearer understanding of how global disruptions interact with local research cultures to shape the future of AI policy research and scientific communication.

## 2. Literature Review

### 2.1 Preprints in Research: Background and Evolution

Preprints are research papers and scholarly articles made available to the public before undergoing a formal peer review (Patel, 2023). Preprints allow researchers to disseminate their findings to the public without the delay of traditional academic publishing. Although preprints date back to the 1960s, they gained popularity in the 1990s with the introduction of an open-access archive known as arXiv (Sarli, 2022). Today, arXiv hosts over two million open-access papers spanning fields such as physics, mathematics, computer science, biology, finance, and engineering (About arXiv, n.d.). Following its success, domain-specific platforms like bioRxiv and medRxiv emerged for life sciences and medicine. Preprints have since transformed scholarly communication by offering speed, greater accessibility, and opportunities for early feedback from the academic community (Klebel et al., 2020). This transformation is significantly relevant in policy-oriented fields such as AI governance, where timely dissemination can directly influence regulatory debates and public discourse.

Prior to the COVID-19 pandemic, preprints were primarily used in fields like physics, computer science, and artificial intelligence (AI). In contrast, medical and biological fields were more hesitant due to concerns about credibility and misinformation of preprints (Vale, 2015). However, the urgency of the pandemic disrupted these norms, as policymakers and researchers required quick access to data on treatments, vaccines, and public health strategies. The slower pace of peer-reviewed publishing could not meet this demand, positioning preprints as the primary method for sharing critical research (Fraser et al., 2021). This shift significantly increased their visibility and acceptance, establishing preprints as a central mode of communication across disciplines (Vale, 2015). As a result, researchers became more comfortable citing and publishing their work through platforms like arXiv, medRxiv, and bioRxiv (Fraser et al., 2021).

### 2.2 Factors Driving Preprint Adoption in AI Policy Research

Recent research claims the growing preference for preprints in AI research represents a structural shift rather than a passing trend. While the pandemic accelerated this change, the rapid pace of innovation in AI and the strong involvement of industry continue to drive preprint adoption (Fraser et al., 2021; Kwon, 2020). Although peer-reviewed journals remain important for quality assurance, preprints now play a critical role in AI research by enabling faster dissemination, broader accessibility, and early impact (Abdill & Blekhman, 2019).

A key factor is the time involved in the peer-review process, which can take up to six months to two years due to multiple rounds of review and editorial stages (Björk & Solomon, 2013). By the time the research paper is published, the field may have already moved on, making the findings outdated (Vale, 2015). Preprints counter this problem by allowing researchers to share work immediately—an essential benefit in a field where advances in algorithms, architectures, and policy frameworks evolve rapidly (Fraser et al., 2021).

Industry influence is another major driver. Tech companies like Google, OpenAI, DeepMind, and Meta have released major reports of breakthroughs such as GPT-3, AlphaFold, and DALL-E on preprints rather than waiting for journal approval (Kwon, 2020). This strategy emphasizes speed, visibility, and influence. Open research practices also help these companies attract talent, foster collaboration, and reinforce leadership in the AI sector (OECD, 2020). In many cases, their innovations are deployed in real-world applications before they ever appear in traditional publications, further reducing reliance on the peer-review model (Heaven, 2018).

### 2.3 Theoretical Framework: The Open Science Movement

The increasing use of preprints in AI policy research is strongly linked to the Open Science movement, which promotes transparency, inclusivity, and accessibility in scientific communication (UNESCO, 2021). Open Science challenges traditional publishing systems by removing barriers such as paywalls and extended peer-review timelines, aiming to make research freely available to the global public (Tennant et al., 2016). Preprint platforms like arXiv, bioRxiv, and medRxiv embody these principles by enabling immediate access to research, encouraging early collaboration, and speeding up knowledge transfer to policy and practice (Bourne et al., 2017). In a fast-moving field like AI, where up-to-date insights are vital for innovation and governance, this open-access model explains why preprints have become integral to research workflows. Adopting these practices allows researchers and institutions to respond more quickly to emerging challenges, support evidence-based policymaking, and foster greater public engagement in science (Fraser et al., 2021). By using the Open Science movement as a theoretical framework, this study connects the growing adoption of preprints to broader systemic changes in scientific communication, providing a foundation for examining how these practices vary across regions and are shaped by different institutional and cultural contexts.

### 2.4 Citation Trends and Regional Differences in AI Policy Research

The way AI policy research is communicated and cited varies significantly across countries, reflecting differences in institutional culture, infrastructure, and policy priorities. The United States, Europe, and South Korea—three leaders in AI development—have each taken unique approaches to balancing preprints and traditional peer-reviewed publishing.

In the United States, preprints have become prominent due to government support, industry leadership, and a strong culture of open research. The Trump administration's Executive Order on Maintaining American Leadership in Artificial Intelligence (2019) encouraged collaboration between academia, industry, and government, with a focus on open-access dissemination (White House, 2019). This policy context supported the alignment between national goals and the speed and visibility offered by preprints. Tech firms like Google, OpenAI, and DeepMind have set a precedent by releasing landmark models like GPT-3, DALL·E, and AlphaFold on platforms like arXiv (Kwon, 2020). This practice has normalized preprint usage in both academic and policy environments. Federal policy further reflects this shift, as documents like the Blueprint for an AI Bill of Rights (2022) and the AI Risk Management Framework (NIST, 2023) cite preprints and open-source materials. These references signal institutional trust and underscore the value of preprints for timely policymaking.

In contrast, Europe has long favored peer-reviewed publications as the standard for research credibility and policy development. This cautious approach reflects a strong commitment to credible scholarly research. However, the COVID-19 crisis pushed many European institutions to adopt preprints out of necessity (Fraser et al., 2021). With peer review too slow to meet urgent needs, preprints became a vital tool for real-time communication. Initiatives like the European Open Science Cloud (EOSC), launched in 2020, accelerated this shift by promoting open-access publication under FAIR principles—Findable, Accessible, Interoperable, and Reusable (Bertelli et al., 2025). EOSC encouraged greater transparency and legitimacy for preprints across EU member states. Nonetheless, the EU continues to rely heavily on peer-reviewed work in official regulatory texts, such as the Artificial Intelligence Act (European Commission, 2021). This dual stance illustrates a transitional phase, where preprints are increasingly used for early access but are not yet considered as a definitive source in formal policy making.

South Korea, meanwhile, adopted a proactive stance toward open access early on. Its National Strategy for Artificial Intelligence (2019) highlighted not only technological leadership but also the democratization of knowledge through open-access publishing (Ministry of Science and ICT, 2019). This early commitment encouraged the use of preprints even before the pandemic. AI policy documents—such as the AI Ethics Guidelines published by the Ministry of Science and ICT and the Korea Information Society Development Institute—explicitly cite preprints, signaling high institutional trust (KISDI, 2021). South Korea's integration of preprints into policy development reflects a broader national strategy that sees openness not as a threat, but as a way to support innovation, agility, and responsiveness in AI governance.

These regional variations highlight how each country balances scientific authority, speed, and access. While the U.S. embraces rapid dissemination led by industry and open science, Europe cautiously adapts within existing academic norms, and South Korea actively incorporates preprints into national strategies. Understanding these differences is essential for evaluating how preprints are shaping the global landscape of AI policy research and informing future models of research governance.

### 2.5 Limitations and Risks of Preprint Adoption

While preprints offer various advantages in speed, accessibility, and early feedback, the growing usage of preprints in scientific discussion and policy development also raises several concerns. One of the main criticisms is the absence of formal peer review, which can compromise the credibility and reliability of a study. Unlike traditional peer-reviewed publications, preprints are not subjected to editorial standards, increasing the risk of methodological flaws, unsupported conclusions, or even misinformation (Vale, 2015; Johansson et al., 2018). This concern is especially important in high stake fields like medicine, public health, and AI governance, where inaccurate information from preprints can cause a significant problem. For example, during the COVID-19 pandemic, preprints were cited widely in the media and political discourse before undergoing any validation, sometimes contributing to the confusion or the spread of unverified claims (Fraser et al., 2021). Additionally, the accessibility of preprints may lead to their misuse by non-expert audiences. Policymakers, journalists, and the public may not always be able distinguish the difference between peer reviewed research and unreviewed preprints, potentially leading to giving excessive credibility to unsubstantiated or unconfirmed results (Klebel et al., 2020).

Despite these challenges, raising awareness about the correct use of preprints, along with clearer labeling practices and post publication peer review initiations may help mitigate these risks. Nonetheless, the integration of preprints in policy and public discourse must be carefully managed and reviewed to ensure that the scientific credibility is not undermined.

## 3. Methodology

### 3.1 Research Design

This study adopts a cross-regional analysis to evaluate how preprint citation trends in AI - policy research have shifted over time in response to two major global events: the COVID-19 pandemic and the release of ChatGPT. This approach was chosen for this study because it allows us to observe the shifts in citation behavior surrounding key events, with a focus on comparing patterns across the United States, Europe, and South Korea.

### 3.2 Data Sources

For this experiment, I collected publicly available bibliometric data from the Web of Science database, using the keyword *"AI Policy"* and applying a location filter for the United States, Europe, and South Korea (See Figure 1). For the European region, this study includes data from a comprehensive list of countries classified geographically and politically as part of Europe. These include countries such as England, Germany, France, Spain, Italy, the Netherlands, and Switzerland, along with others such as Sweden, Belgium, Norway, Finland, Portugal, Poland, Greece, Austria, Ireland, Romania, Denmark, and Hungary. Additionally, smaller countries like Estonia, Latvia, Lithuania, Bulgaria, Croatia, Slovenia, Slovakia, Ukraine, North Macedonia, Serbia, Albania, Moldova, Bosnia and Herzegovina, Montenegro, Kosovo, Iceland, Malta, Cyprus, Luxembourg, Liechtenstein, Monaco, San Marino, Scotland, Wales, and Northern Ireland were also considered. The analysis focused on the number of AI policy-related papers that cited arXiv preprints annually between 2018 to 2024, offering a consistent indicator of preprint adoption in policy-focused AI research.

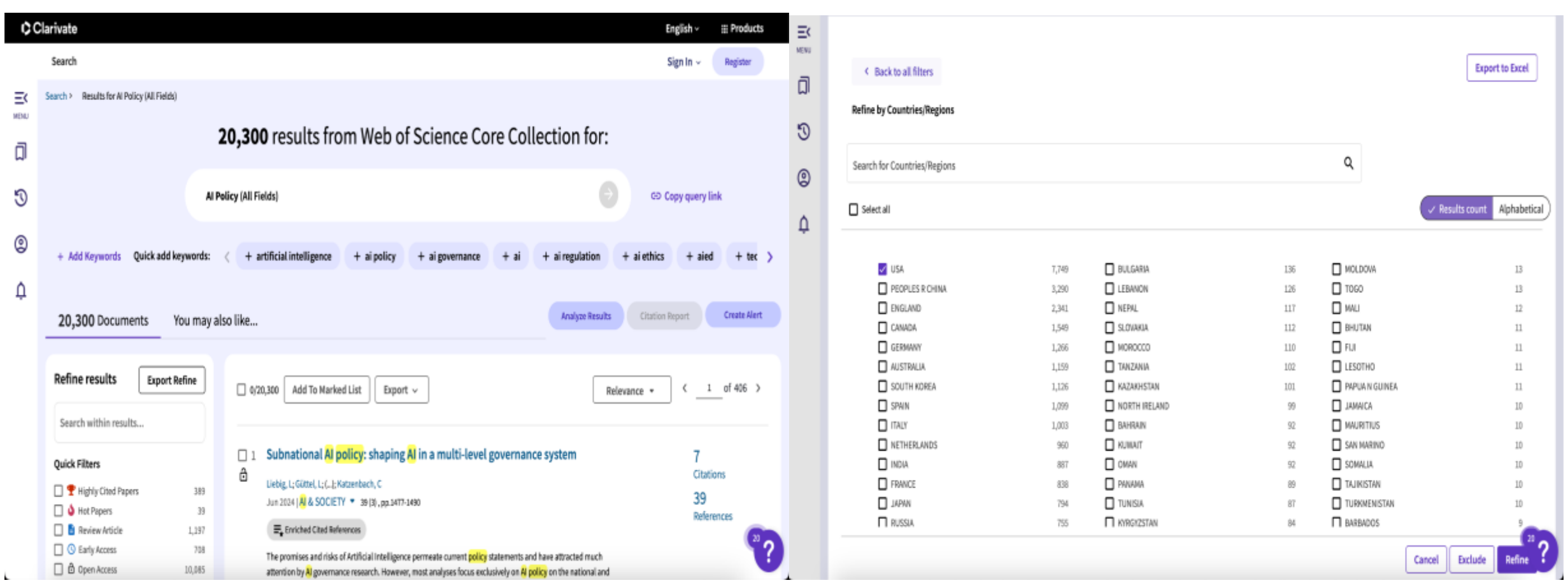

Figure 1: Keyword search of "AI Policy" in Web of Science (Left) / Location filter of United States with the same keyword in Web of Science (Right)

### 3.3 Data Collection

After the data was searched and filtered, data were collected by performing structure queries within the Web of Science interface. For the publications that cited arXiv preprints during the year 2018 to 2024, the data of full records and cited references was exported into a BibTeX format. Due to Web of Science's export limit of 500 records per session, the exporting process was repeated until all the data files were exported. The final dataset resulted in:

- 7,749 records for the United States
- 13,918 records for Europe
- 1,126 records for South Korea

All records that cited an arXiv preprints were retained regardless of whether they citing paper was itself a preprint or a pre-reviewed article.

### 3.4 Data Analysis

The data analysis for this study was conducted using a cross-regional comparison to evaluate the changes in preprint citation trends over time. This approach allowed the study to capture both the temporal dynamics of preprint adoption and the regional variations in response to two key intervention points:

1. the onset of the COVID-19 pandemic (early 2020)
2. the release of ChatGPT (late 2022)

As shown in Figure 2, these events were selected based on their significant influence on research dissemination practices and public discourse surrounding AI. The analysis focused on comparing the total number of the keywords “AI” and “policy” that cited arXiv preprints before and after each event across the United States, Europe, and South Korea. By observing shifts in publication trends across regions, the study aims to reflect on how global disruptions shaped the evolving relationship between open science practices and AI policy development within different institutional and cultural contexts.

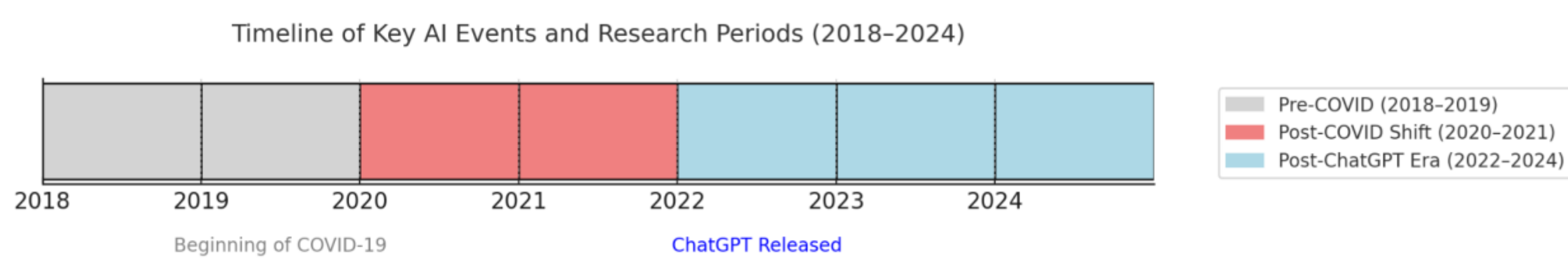

Figure 2: Timeline of Key AI Events and Research Periods(2018-2024)

### 3.5 Limitations

One of the key limitations of this study lies in the data collection process. Although a comparative analysis using multiple scholarly databases—such as Google Scholar, ResearchGate, and the Directory of Open Access Journals (DOAJ)—was initially considered to enhance the scope and reliability of the citation data, this approach was not possible due to several constraints. Many platforms impose daily export limits, restrict full data access behind paywalls, or require administrative approval for large-scale data extraction.

Similar challenges were encountered when attempting to extract regional data. For South Korea, searches were conducted using the same keywords on local scholarly platforms such as RISS, KCI, and DBPIA. For Europe, the same effort was made on platforms like OpenAIRE, CORE, and HAL. However, these attempts were not feasible by the same access restrictions, along with additional language barriers that complicated the search and retrieval processes.

Due to these limitations and the short timeframe available for this research, this study ultimately relied only on the data retrieved from the Web of Science database. Specifically, the dataset of research publications in Web of Science that cited arXiv preprints.While Web of Science is a reputable and widely used source for bibliometric analysis, relying on a single database may have limited the comprehensiveness and generalizability of the findings, particularly in capturing region-specific publication trends.

This study also relies on annual publication data, which limits the granularity of trend detection and prevents month-to-month statistical comparisons. Due to the absence of descriptive publication data, no statistical tests were applied.

Additionally, due to time constraints, the writing process—particularly for the literature review and structural organization of the paper—was supported with the use of ChatGPT. The tool was employed to help improve clarity, structure, and consistency across sections. While the research, data collection/analysis, and interpretations are entirely original, this acknowledgment is made for the sake of transparency of this study.

## 4. Results

### 4.1 Regional Trends Over Time (2018 - 2024)

To understand how preprint usage has evolved across global AI research communities, this study analyzed bibliographic data collected from the Web of Science database. The dataset includes research publications from Web of Science that contains the keywords “AI” and “policy” and that cites at least one arXiv preprint references in their paper. Something to note is that this analysis does not examine preprints that are hosted on arXiv directly, nor does it focus on official policy documents or government publications. Rather, the goal is to track citation behavior within formal academic publications and explore how different regions integrate arXiv preprints into scholarly communication on AI-related policy topics.

In addition to tracking citations of arXiv preprints, this study also analyzed overall peer-reviewed publication trends based on papers indexed in Web of Science using the keywords “AI” and “policy.” (See Figure 3). This baseline enables a more accurate interpretation of whether the observed changes in preprint citation behavior are part of broader increases in publication volume, or represent a distinct shift toward open science dissemination.

The time frame of 2018 to 2024 was selected to capture trends before and after two major global events: the COVID-19 pandemic (2020) and the release of ChatGPT (late 2022). By examining the annual change in the number of arXiv citations within this time frame, this section highlights the regional difference in how preprints have been adopted across the United States, Europe, and South Korea.

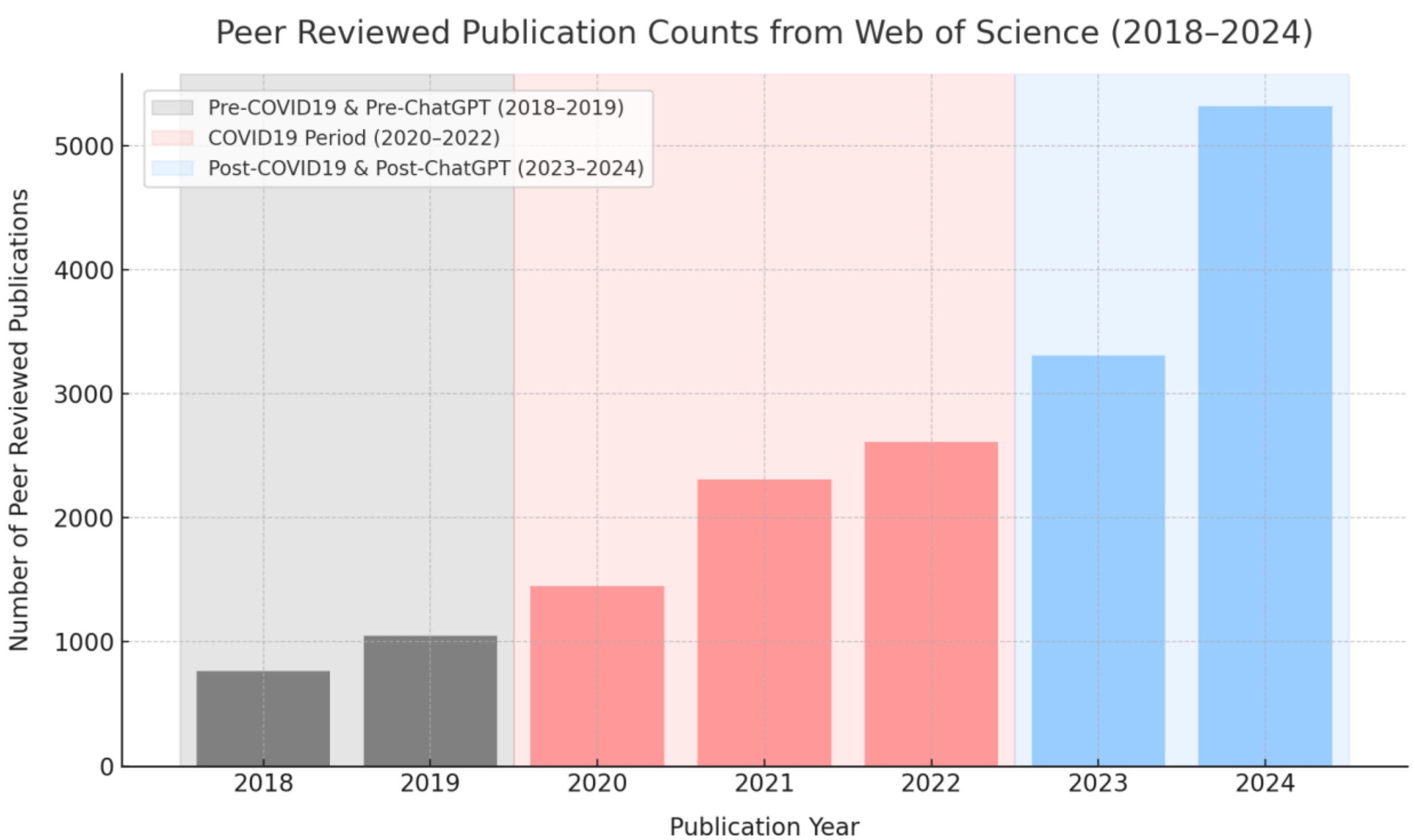

Figure 3: Number of Peer Reviewed Publications in Web of Science Over the Years

### 4.1.1 United States

The number of Web of Science indexed publications in the United States containing the keywords “AI” and “policy” that cited at least on arXiv preprints increased significantly between the years 2018 and 2024. As shown in Figure 4, arXiv preprints citations rose modestly from 2018 to 2019 but then dipped slightly in 2020 – the year the COVID-19 pandemic began to disrupt research publishing globally. This short-term decline may reflect initial uncertainty and delays within academic publishing as institutions adapted to work remotely and journal operations were slowed down.

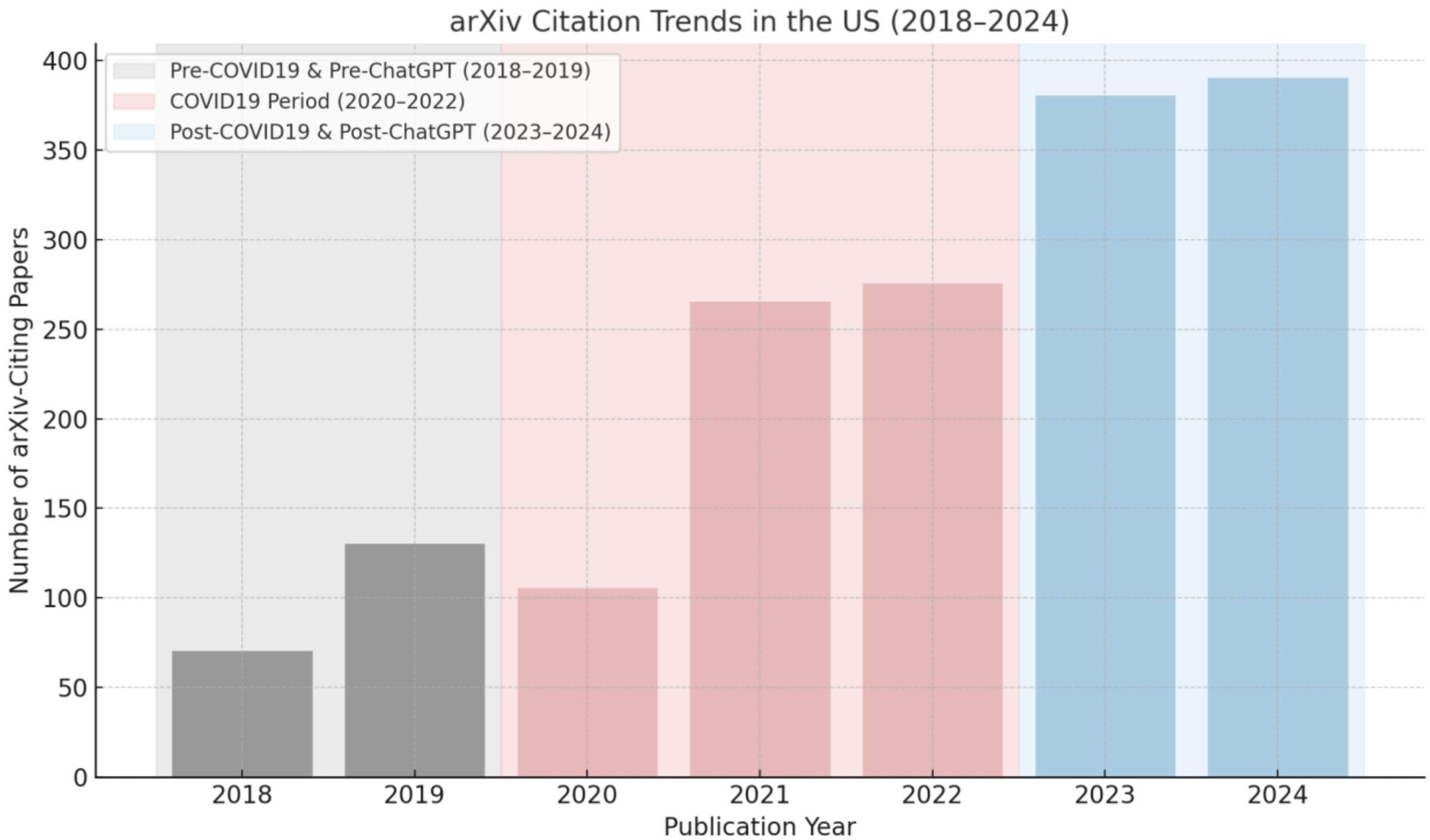

Figure 4: Number of arXiv-Referenced Papers Published in the United States Over the Years (Pre/Post COVID-19) (Web of Science Analysis)

Following this brief decline, there was a sharp and sustained increase in preprint citations from 2021 onward. The numbers of papers citing arXiv nearly doubled by 2022 compared to pre-pandemic levels, and grew even more rapidly after the release of ChatGPT in late 2022. This trajectory suggests that researchers increasingly turned to preprints as a faster and more accessible means of disseminating information during a period when traditional peer-reviewers faced significant delays in processing.

The delayed boost in U.S. citation indicates a broader shift across regions, where preprints gained legitimacy. This pattern may also reflect a growing acceptance of arXiv within AI-related policy research, particularly as industry and academic researchers sought faster sources to influence public and regulatory discussions. As further shown in Figure 5, arXiv was the most frequently cited source in AI policy publications, highlighting the growing reliance on preprints in the U.S. research ecosystem.

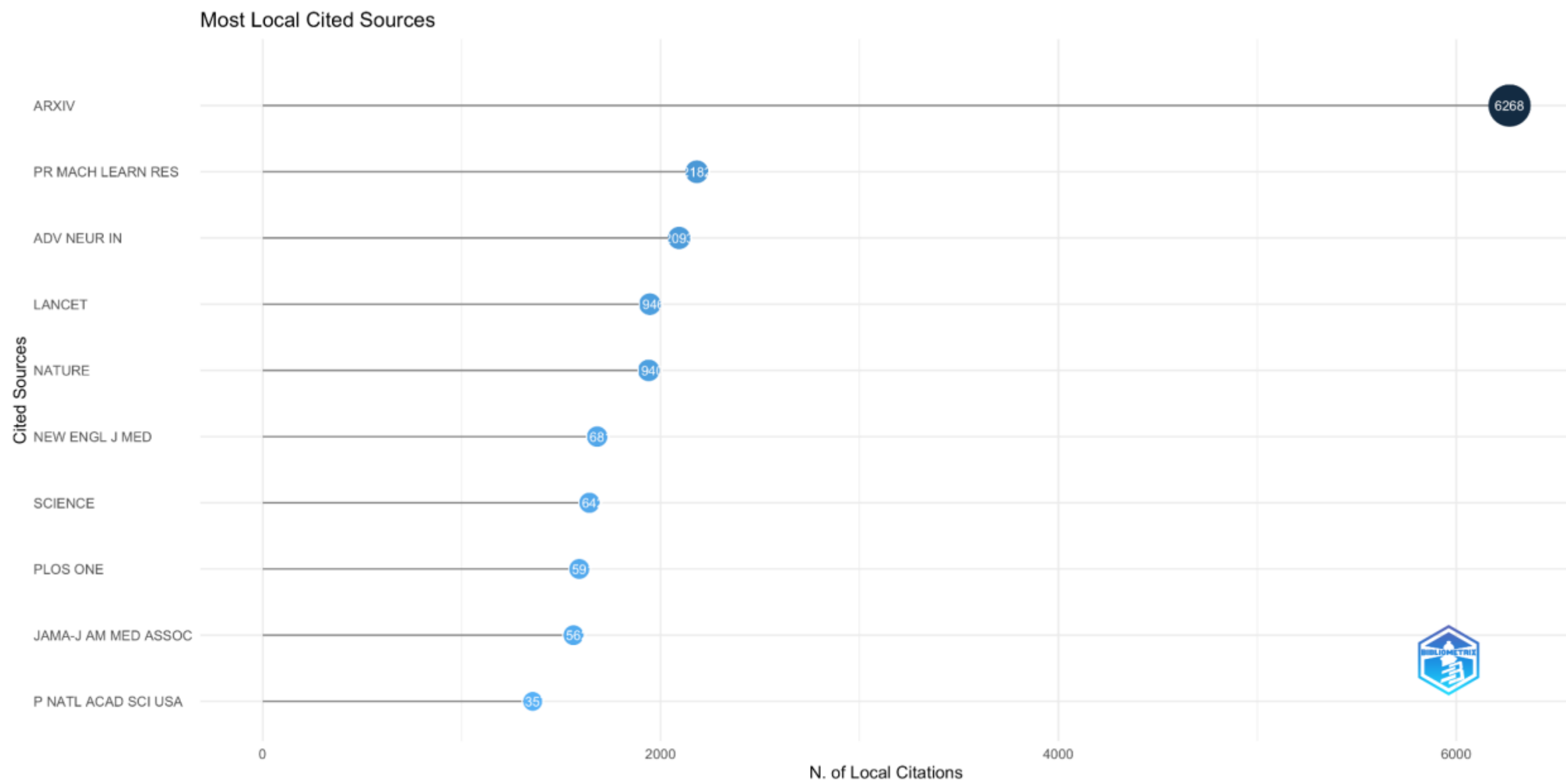

Figure 5: Top Locally Cited Sources for “AI” and “Policy” related research in the United States (Web of Science Analysis)

### 4.1.2 Europe

In Europe, a strong upward trend in preprint citations was observed, though the rate of change was more gradual than the United States. As shown in Figure 6, the number of Web of Science-indexed publications containing the keywords “AI” and “policy” that cited arXiv preprints tripled compared to pre-pandemic levels. While there were no notable changes around 2020, the growth accelerated significantly after the release of ChatGPT in late 2022.

This steady trajectory suggests that European researchers may have turned to preprints in response to the same global disruptions – such as delays in peer reviewed publishing during the pandemic - but did so within a more cautious and institutionally guided framework. The absence of a sharp increase or decrease reflects a slower institutional adoption to open-access practices, rather than a sudden shift.

The increase in citations may also reflect delays in traditional journal publishing. As peer review took longer during the pandemic, preprints became an effective alternative for disseminating findings. Europe’s growing use of preprints during this time also aligns with broader Open Science initiatives, which encouraged transparency and accessibility even during periods of disruption.

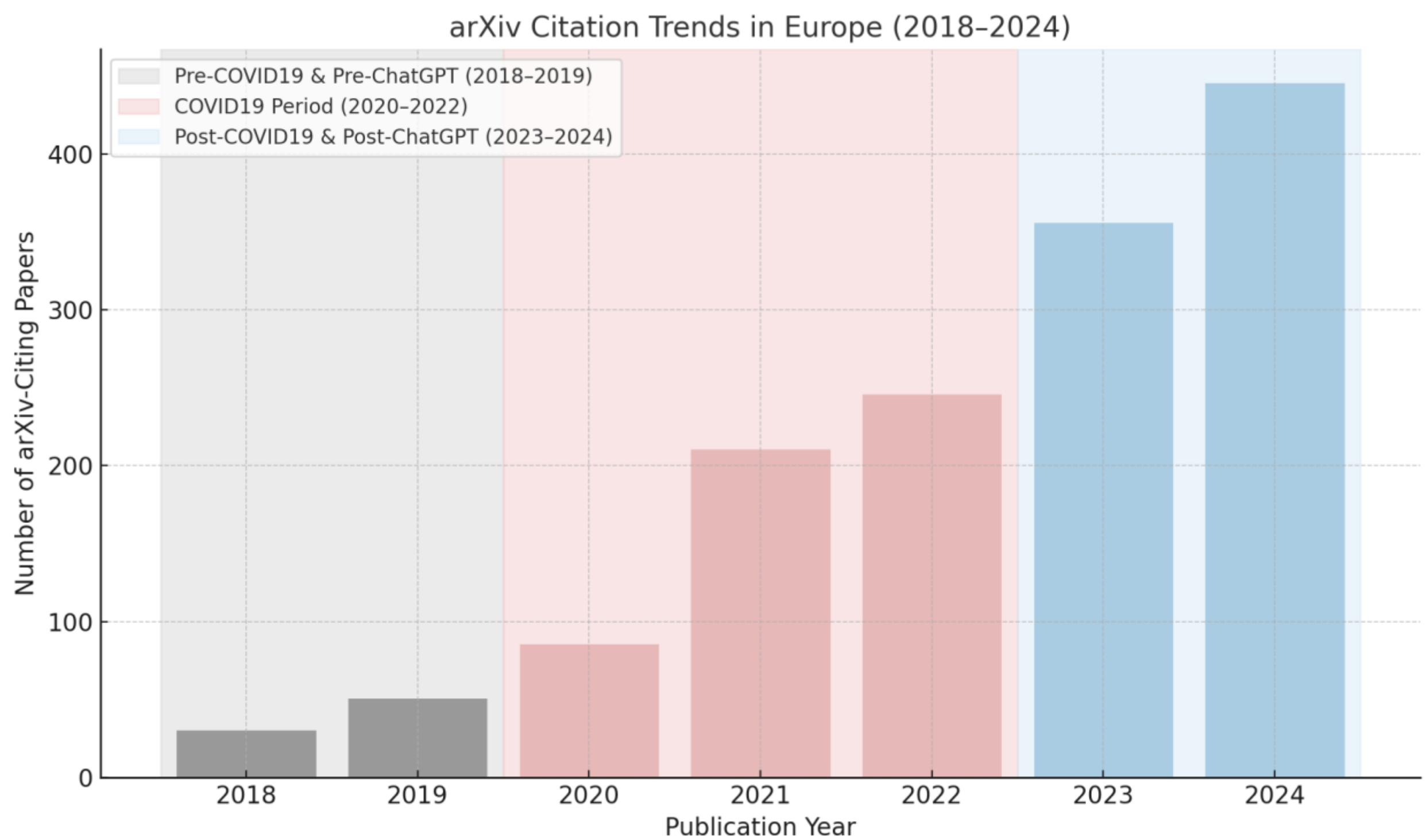

Figure 6:Number of arXiv-Referenced Papers Published in Europe Over the Years (Pre/Post COVID-19) (Web of Science Analysis)

Analysis of locally cited sources shows that arXiv has also become the most cited source for AI and policy related research across European publications (see Figure 7). This reflects a transitional phase in Europe, where preprints are increasingly being accepted to research.

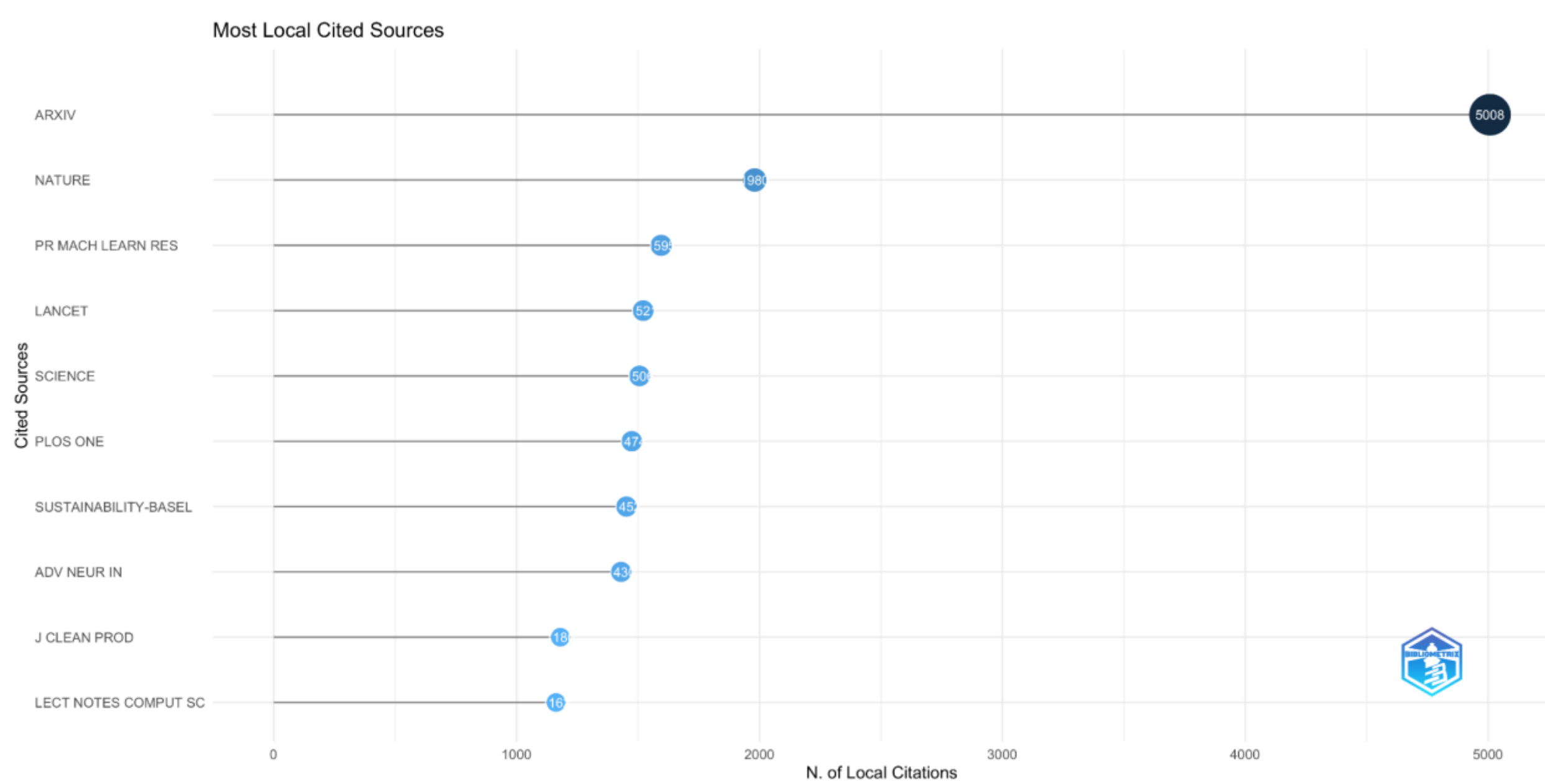

Figure 7: Top Locally Cited Sources for “AI” and “Policy” related research in Europe (Web of Science Analysis)

### 4.1.3 South Korea

South Korea exhibited a different pattern compared to the United States and Europe. The number of Web of Science-indexed publications from South Korea containing the keywords *"AI"* and *"policy"* that cited at least one arXiv preprint showed consistent growth between 2018 and 2024, but with less dramatic fluctuations around major global events. As shown in Figure 8, citation levels rose gradually from 2018 to 2019, slightly decreased in 2020, and then resumed a steady upward trend through 2024.

The slight decrease in 2020 may reflect a delayed impact of the COVID-19 pandemic in South Korea's research infrastructure, compared to earlier disruptions observed in the U.S. The timeline of national lockdowns and institutional adjustments differed by region, and this may have influenced when researchers faced problems with traditional publishing methods.

Unlike the U.S. and Europe, South Korea did not experience a post-ChatGPT spike in preprint citation trends. Instead, its preprint citation trend followed a consistent upward linear trend indicating that open-access dissemination through arXiv may have already been integrated in the institution before the external shocks occurred. This stability supports the idea that South Korea had adopted preprints as part of a broader national strategy to promote open science, as highlighted in its National Strategy for Artificial Intelligence (Ministry of Science and ICT, 2019). Rather than reacting to external disruptions, the growth of preprint citations in South Korea appears to reflect a long term institutional commitment to accessible and quick research dissemination.

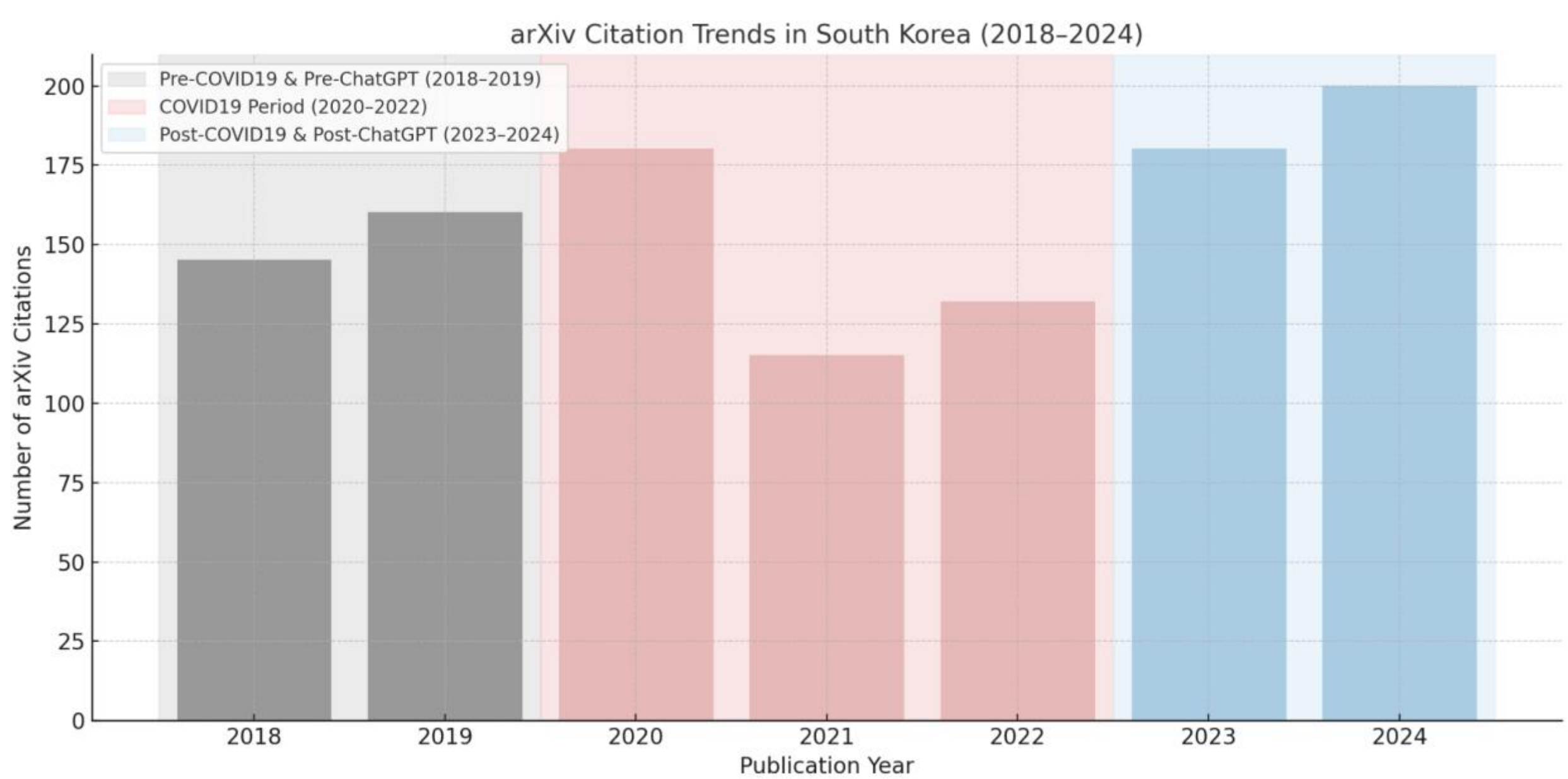

Figure 8: Number of arXiv-Referenced Papers Published in South Korea Over the Years (Pre/Post COVID-19) (Web of Science Analysis)

Analysis of locally cited sources also confirms that arXiv is the dominant cited platform for AI and policy related research in South Korea (see Figure 9), emphasizing the country's strong institutional trust in open-access research models.

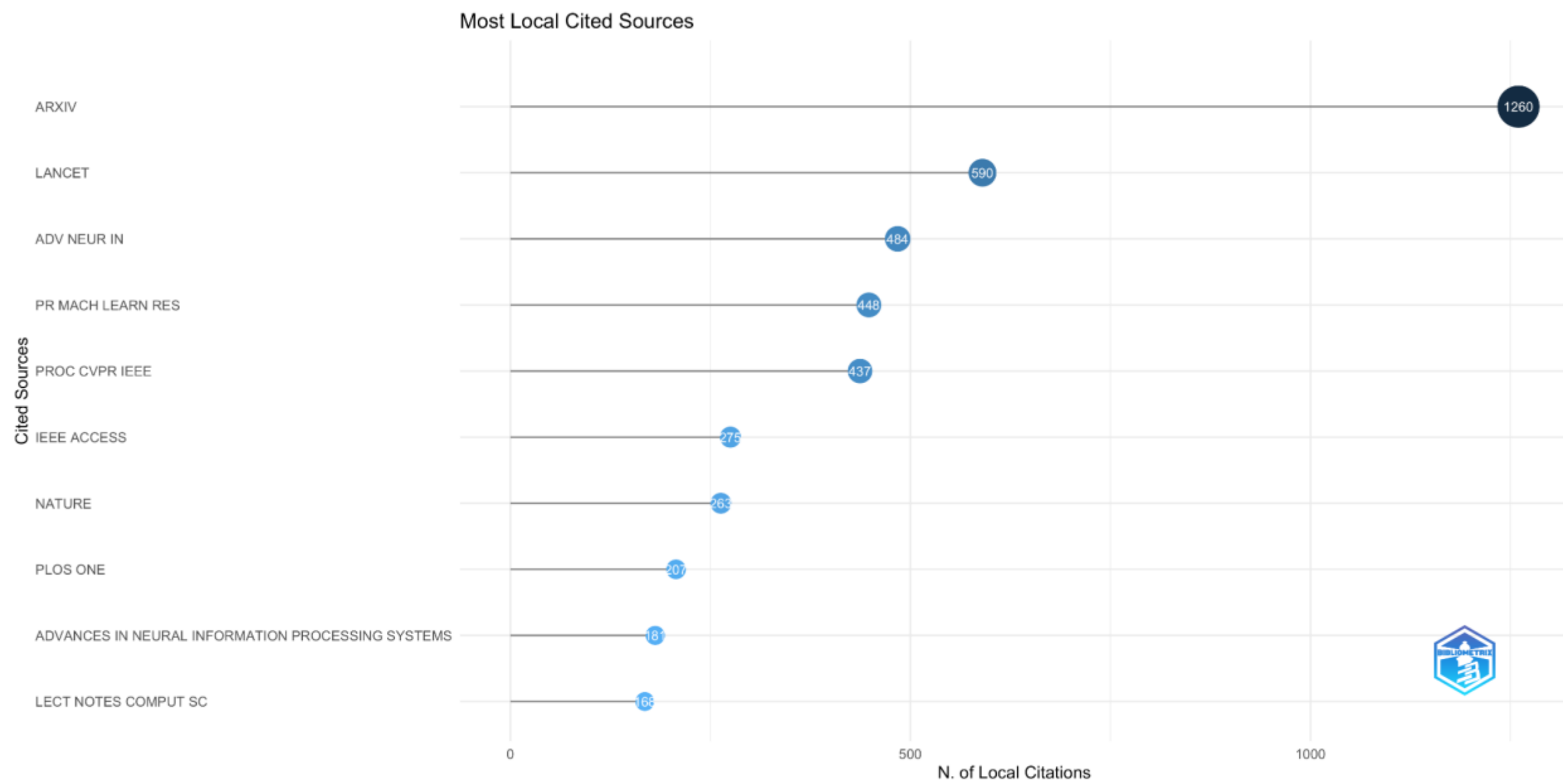

Figure 9: Top Locally Cited Sources for “AI” and “Policy” related research in South Korea (Web of Science Analysis)

## 4.2 Regional Trends Over Time

The decision to examine regional trends over time was made to evaluate whether the adoption of open science practices, such as citing arXiv preprints, was influenced by two major external interventions - the onset of the COVID-19 pandemic and the release of ChatGPT - or whether if it reflected distinct regional dynamics. Specifically, this analysis seeks to determine whether the acceleration of preprint citations represents a unified global trend or reveals unique trends shaped by local research cultures, institutional environments, and policy frameworks. By comparing the United States, Europe, and South Korea, this study aims to assess both the global influence of these disruptive events and the extent to which regional differences lead to the adoption of open science norms. Understanding these patterns is crucial for developing future strategies that promote responsible and equitable research practices across diverse international contexts.

As shown in Figure 9, the regional trend analysis reveals notable differences across the United States, Europe, and South Korea. In the United States, citations of preprints grew steadily from approximately 250 papers in 2018 to about 400 papers in 2020, with further growth to around 620 papers by 2022. Following the release of ChatGPT, the U.S. showed a sharp increase in citations, which may reflect a growing uptake of open science norms in response to technological innovation and heightened public interest in AI. In Europe, the trend was similar; citations of preprints rose from approximately 190 papers in 2018 to 380 in 2020, and continued steadily to around 660 by 2022. After the release of ChatGPT, Europe's citation counts rose even more sharply, totaling approximately 1200 papers by 2024. This suggests that European researchers not only responded to global innovations but also sustained a faster growth trajectory than the U.S., potentially due to institutional efforts to promote open-access norms across member states. In contrast, South Korea displayed a more gradual but consistent growth pattern. Citations of preprints increased from approximately 15 papers in 2018 to 50 in 2020, and continued rising steadily to around 130 papers by 2022. After ChatGPT's release, South Korea's citation numbers grew modestly, reaching approximately 220 papers by 2024. While South Korea's overall citation volume remained lower compared to the United States and Europe, this difference is likely attributable to its smaller research base and lower overall publication volume, rather than a lower rate of preprint adoption relative to national output. Its consistent upward trend reflects a growing integration of open science practices within its research community.

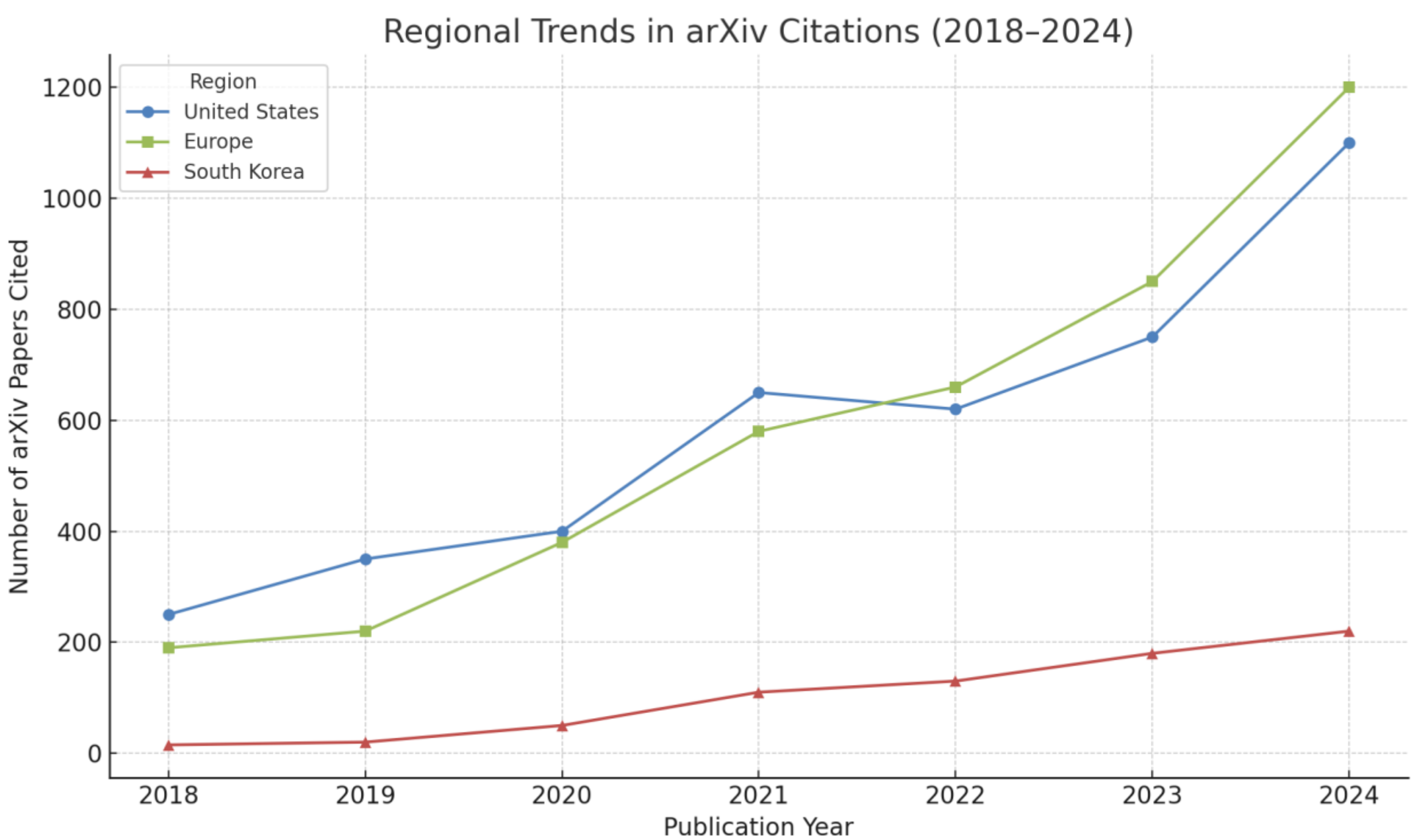

Figure 9: Regional Trends in arXiv citations for U.S., Europe, and South Korea

## 5. Discussion

### 5.1 Summary of Main Findings

This study found that citations of arXiv preprints in AI-related research publications (identified through the keywords "AI" and "policy") have accelerated globally, though the magnitude and pattern of change vary significantly across regions. In the United States, the onset of the COVID-19 pandemic corresponded with a substantial increase in citations of arXiv preprints, followed by an even sharper rise after the release of ChatGPT. Europe exhibited a similar trajectory, but with a more gradual and sustained upward trend, ultimately reaching citation levels comparable to or exceeding those of the United States by 2024. In contrast, South Korea displayed a unique pattern of steady, uninterrupted growth in preprint citations throughout the time period. These regional differences highlight the interaction between global scientific disruptions and local research cultures in shaping the adoption of open science practices.

### 5.2 Interpretation of Regional Differences

The observed regional variations can be interpreted through the lens of institutional, cultural, and policy-driven factors. In the United States, the sharp rises following major global events are consistent with a highly responsive research ecosystem, characterized by strong industry leadership and governmental support for rapid knowledge dissemination. The steep increase after ChatGPT's release suggests a heightened reliance on preprints as a mechanism for maintaining leadership in AI innovation.

In Europe, the pattern of steady growth and stabilization reflects a research environment that values both speed and credibility. While the COVID-19 pandemic and ChatGPT influenced preprint adoption, European researchers appear to have integrated preprints into existing academic frameworks more cautiously, balancing open science initiatives with traditional peer-review standards.

On the other hand, South Korea's consistent upward trajectory indicates an early and proactive embrace of open science principles, driven by national strategies that emphasize knowledge democratization. The absence of sudden spikes suggests that preprint adoption was already institutionalized before external global events, positioning South Korea as a stable adopter of open-access research models.

### 5.3 Implications for AI Policy Research

The increasing reliance on preprints is reshaping the communication and dissemination of AI policy research. Preprints enable faster sharing of research outputs, making them more accessible to policymakers, researchers, and the broader public. However, the regional differences observed in this study highlight that the transition to open science is not uniform.

Effective AI governance strategies must recognize these differences and tailor policies that address regional institutional cultures, levels of technological infrastructure, and historical attitudes toward peer review and open-access practices. Failing to do so could reinforce existing disparities in research accessibility and influence the global distribution of scientific knowledge and leadership in AI policy development.

### 5.4 Limitations

Several limitations for this study must be acknowledged. First, the 2025 data only covers the first four months of the year, which may exaggerate the appearance of a post-ChatGPT decline in the data. Second, the study relied solely on the Web of Science database, while reputable Web of Science data may not capture the full spectrum of preprint adoption behaviors, especially in non-English or regional publications. Finally, this study focused on citation behavior and did not assess the content or quality of either the cited papers or preprints themselves.

### 5.5 Suggestions for Future Research

Future studies should extend the data collection period beyond 2025 to better capture long-term post-ChatGPT trends and evaluate whether the observed peaks and stabilization in preprint citations are sustained or temporary. In addition to expanding the temporal scope, a more in-depth analysis would be beneficial. For example, future research could incorporate multiple bibliometric databases such as Google Scholar, CORE, and national preprint repositories to enhance the reliability and generalizability of the results. We could also download, open, and sort individual preprints by month of publication while gathering data from these bibliometric databases. This would allow for a detailed breakdown of annual publication volumes into monthly ratios, offering a deeper insight into the dynamics of preprint adoption around key intervention points such as the COVID-19 pandemic and the release of ChatGPT. Finally, gathering quantitative data with qualitative research, such as interviews with researchers, policy advisors, and journal editors, could provide valuable context regarding the motivations, attitudes, and institutional barriers influencing preprint usage across different regions.

## 6. Conclusion

This study examined how two major global events—the COVID-19 pandemic and the release of ChatGPT—impacted the citation of preprints in AI-related research publications across the United States, Europe, and South Korea. The analysis was based on bibliometric data retrieved from the Web of Science, filtered by the keywords “AI” and “policy,” and focused specifically on publications that cited arXiv preprints. The findings reveal that while preprint citation increased globally, the magnitude and nature of change varied significantly by region.

The United States exhibited sharp, event-driven shifts; Europe demonstrated a gradual yet accelerating trajectory; and South Korea maintained a steady and consistent rise with minimal disruption.

These results underscore the importance of regional research cultures, institutional frameworks, and policy environments in shaping the global landscape of open science practices. Understanding these dynamics is crucial for designing governance strategies that promote transparency, inclusivity, and equitable participation in scientific communication. As preprints continue to play an increasingly central role in the dissemination of AI-related policy research, future governance models must accommodate regional variability while upholding the principles of open science. Expanding longitudinal analyses and incorporating qualitative insights will be essential for tracking the evolving relationship between preprints, policy, and innovation in the global AI research ecosystem.

## References

Abdill, R. J., & Blekhman, R. (2019). Tracking the popularity and outcomes of all bioRxiv preprints. *eLife, 8*, e45133. https://doi.org/10.7554/eLife.45133

About arXiv – arXiv info. (n.d.). https://info.arxiv.org/about/index.html

Bertelli, A., Acciai, M., & Rossi, G. (2025). The European open science cloud as a common good Potentials and limitations of this endeavour. *Open Research Europe*, *5*, 19. https://doi.org/10.12688/openreseurope.19170.1

Björk, B. C., & Solomon, D. (2013). The publishing delay in scholarly peer-reviewed journals. *Journal of Informetrics, 7*(4), 914–923. https://doi.org/10.1016/j.joi.2013.09.001

Bourne, P. E., Polka, J. K., Vale, R. D., & Kiley, R. (2017). Ten simple rules to consider regarding preprint submission. *PLOS Computational Biology, 13*(5), e1005473. https://doi.org/10.1371/journal.pcbi.1005473

European Commission. (2021). *Proposal for a Regulation laying down harmonised rules on artificial intelligence (Artificial Intelligence Act)*. https://eur-lex.europa.eu/legal-content/EN/TXT/?uri=CELEX:52021PC0206

Fraser, N., Brierley, L., Dey, G., Polka, J. K., Pálfy, M., Nanni, F., & Coates, J. A. (2021). Preprinting the COVID-19 pandemic. *bioRxiv.* https://doi.org/10.1101/2020.05.22.111294

Heaven, D. (2018). Why self-archiving is good for science. *Nature, 561*(7721), 445–447. https://doi.org/10.1038/d41586-018-06742-8

Johansson, M. A., Reich, N. G., Meyers, L. A., & Lipsitch, M. (2018). Preprints: An underutilized mechanism to accelerate outbreak science. *PLOS Medicine, 15*(4), e1002549. https://doi.org/10.1371/journal.pmed.1002549

Kelly, J., Sadeghieh, T., & Adeli, K. (2014). Peer review in scientific publications: Benefits, critiques, & a survival guide. *EJIFCC, 25*(3), 227–243. https://www.ncbi.nlm.nih.gov/pmc/articles/PMC4975196/

Klebel, T., Reichmann, S., Polka, J., McDowell, G., Penfold, N., Hindle, S., & Ross-Hellauer, T. (2020). Peer review and preprint policies are unclear at most major journals. *PLOS ONE, 15*(10), e0239518. https://doi.org/10.1371/journal.pone.0239518

Korea Information Society Development Institute (KISDI). (2021). *The National Guidelines for AI Ethics*. https://ai.kisdi.re.kr/eng/main/contents.do?menuNo=500011

Kwon, D. (2020). How swamped preprint servers are blocking bad coronavirus research. *Nature.* https://doi.org/10.1038/d41586-020-01394-6

Ministry of Science and ICT. (2019). *National Strategy for Artificial Intelligence*. Republic of Korea.

https://www.msit.go.kr/bbs/view.do?bbsSeqNo=46&mId=10&mPid=9&nttSeqNo=9&sCode=eng
National Institute of Standards and Technology. (2023). *AI risk management framework (AI RMF 1.0)*. https://www.nist.gov/itl/ai-risk-management-framework
National Academies of Sciences, Engineering, and Medicine. (2019). *The growing role of artificial intelligence in scientific discovery.* The National Academies Press. https://doi.org/10.17226/25384
OECD. (2020). *OECD AI policy observatory: COVID-19 and AI.* https://oecd.ai/en/wonk/covid-19-and-ai
Patel, S. (2023, March 27). The pros and cons of preprints. *MDPI Blog*. https://blog.mdpi.com/2023/03/27/preprints-pros-cons/
Sarli, C. (2022, April 22). Preprints: A brief history and recommendations from the NIH for authors citing preprints. *Becker Medical Library.* https://becker.wustl.edu/news/preprints/
Tennant, J. P., Waldner, F., Jacques, D. C., Masuzzo, P., Collister, L. B., & Hartgerink, C. H. J. (2016). The academic, economic and societal impacts of Open Access: An evidence-based review. *F1000Research, 5*, 632. https://doi.org/10.12688/f1000research.8460.3
UNESCO. (2021). *UNESCO recommendation on open science*. United Nations Educational, Scientific and Cultural Organization. https://unesdoc.unesco.org/ark:/48223/pf0000379949
Vale, R. D. (2015). Accelerating scientific publication in biology. *Proceedings of the National Academy of Sciences, 112*(44), 13439–13446. https://doi.org/10.1073/pnas.1511912112
White House. (2019). *Executive order on maintaining American leadership in artificial intelligence.* https://trumpwhitehouse.archives.gov/presidential-actions/executive-order-maintaining-american-leadership-artificial-intelligence/
White House Office of Science and Technology Policy. (2022). *Blueprint for an AI bill of rights: Making automated systems work for the American people.* https://www.whitehouse.gov/ostp/ai-bill-of-rights/